# Magneto-transport measurements on $Nd_{1.83}Ce_{0.17}CuO_{4\pm\delta}$ thin films

A. Guarino[a,b], A. Leo[a,b], F. Avitabile[a], N. Martucciello[b], A. Avella[a,b], G. Grimaldi[b],
A. Romano[a,b], P. Romano[c,b], A. Nigro[a,b]

[a] Physics Department "E R Caianiello", University of Salerno, I-84084 Fisciano (SA), Italy
[b] CNR-SPIN Salerno, University of Salerno, I-84084  Fisciano (SA), Italy.
[c] Science and Technology Department, University of Sannio, I-82100 Benevento, Italy

*Abstract*— $Nd_{2-x}Ce_xCuO_{4\pm\delta}$ (NCCO) epitaxial thin films have been deposited on (100) $SrTiO_3$ substrates by DC sputtering technique in different atmosphere. The as-grown samples show different dependence of the in-plane resistivity at low temperature, when they are grown in pure argon atmosphere or in oxygen. Moreover, an unusual behaviour is also found when transport takes place in the presence of an external magnetic field.  It is commonly accepted that the higher anisotropic properties of NCCO crystalline cell with respect to the hole doped YBCO and LSCO and the electric conduction mainly confined in the $CuO_2$ plane, strongly support the two-dimensional (2D) character of the current transport in this system. Results on the temperature dependence of the resistance, as well as on the magnetoresistance and the Hall coefficient, obtained on epitaxial NCCO thin films in the over-doped region (x ≥ 0.15) of the phase diagram are presented and discussed.

## I. Introduction

Most of the recent works on electron-doped high-$T_c$ superconductors (HTSC) are devoted to the understanding of the electronic structure and of the wide spectrum of phenomena that characterize the complex phase diagram of  these compounds [1-3]. Only few studies have been dedicated to the realization of devices based on electron-doped HTSC cuprates, due to the hard control of the doping elements in these compounds [4, 5]. In $RE_{2-x}Ce_xCuO_{4\pm\delta}$ (RE=Pr, Nd, Sm,La..) superconductors, the doping mechanism seems to be much more complicated than in the hole-doped case, owing to the unclear role of the oxygen atoms in the dynamic of the reduction process, which seems to be essential to observe the superconductivity [1, 6]. This leaves many issues still under debate, such as, for instance, in the case of the multilayer growth combining electron- and hole-doped compounds [5]. In this case the greatest difficulty is to find the right equilibrium between the oxygen reduction, needed to achieve the superconductivity in the electron-doped layers, and the oxygenation at high pressure required by the hole-doped ones. Recently, there was a big increase in the research activity on the growth and characterization of $RE_{2-x}Ce_xCuO_4$ thin films, with RE = Pr, Nd and La [7-11]. The key element is always the role of the oxygen atoms in the reduction process, since the thermal treatments to obtain the superconducting transition often vary with the method of growth [1]. The strong influence of the cerium, as well as of the oxygen doping in the electron-doped compounds, is reflected in their physical properties, from the crystalline structure [7, 9] to the x-ray photoemission spectroscopic analysis [12, 13] and the magneto-transport behavior [10, 14].
In this work, we analyze the magneto-transport properties and their correlation with the structural features of over-doped $Nd_{1.83}Ce_{0.17}CuO_{4\pm\delta}$ films, grown by dc sputtering technique with different content of oxygen. The choice of the over-doped regime is related to the possibility to investigate the change of sign of the Hall coefficient as a function of the temperature [15], as predicted by the two-carrier scenario, widely accepted for the interpretation of this behavior [10, 14]. The role of the oxygen content is analyzed on as-grown samples deposited by changing the mixed oxygen atmosphere [8] and on ex-situ thermally treated superconducting films [16]. The magnetoresistance (MR) results will be discussed and linked to the resistance versus temperature (*R-T*) measurements in order to search for an interpretation of the upturn in the *R-T* characteristics at low temperature [17-19]. We propose that the 2D localization induced by the disorder due to the non-stoichiometric character of the oxygen content is at the origin of the insulating state developing at low temperatures [11, 20]. Finally, the Hall coefficient ($R_H$) is evaluated for as-grown as well as for superconducting films.

## II. EXPERIMENTAL

DC sputtering technique was optimized to grow $Nd_{2-x}Ce_xCuO_{4\pm\delta}$ (NCCO) films on (100) $SrTiO_3$ (STO) substrates by using a single stoichiometric target as a sputtering source in an on-axis configuration with the substrate [8]. The films, 100 nm thick, can be grown in pure argon (Ar) or in a mixed atmosphere of Ar and oxygen ($O_2$), with ratio $O_2/Ar \sim 1\%$ or $O_2/Ar \gg 1\%$, at a total pressure of 1.7 mbar and substrate temperature of about 850 °C. Two kinds of as-grown non-superconducting films were obtained, one containing only oxygen from the target or from the mixed atmosphere with low oxygen pressure (here named *a*-type), the other one enclosing also atoms from the mixed atmosphere with higher oxygen pressure (*b*-type). In both cases, the superconductivity of these films was observed after a suitable combination of thermal treatments in flowing argon at a temperature of about 600 °C and 900 °C. The morphology, phase composition and purity of the samples were inspected by high-resolution x-ray diffraction and scanning electron microscopy combined with wavelength dispersive spectroscopy (WDS) [8]. The magneto-transport properties were investigated by using a standard four-probe technique in a Cryogenic variable temperature cryogen-free system, equipped with a variable temperature insert and a superconducting magnet. Hall measurements were performed in the same system up to 9 T in the range of temperature between 1.6 K and 300 K. The samples were patterned in a standard Hall bar geometry (100 μm wide and 1 mm long strips) using standard photolithography and wet etching in a 1% solution of $H_3PO_4$ in pure $H_2O$.

## III. RESULT AND DISCUSSION

Well-oriented films were obtained without spurious phases with a cerium content of 0.17. The main difference between the above mentioned *a*-type and *b*-type samples is the oxygen content. In fact, it was observed a clear distinction between the two kinds of as-grown films from the x-ray diffraction analysis. θ-2θ measurements around the (004) reflection of the NCCO phase were performed on as-grown *a*-type and *b*-type films and, after the suitable thermal treatments, on the corresponding superconducting films. The analysis of the (00ℓ) reflections allows to obtain directly the *c*-axis lattice parameter from the Bragg law, $2d\sin\theta = n\lambda$, being $d \equiv c$ in this particular case, with $\lambda = 1.54056$ Å, *n* the number of the reflection and θ the half of the 2θ peak position. Fig. 1 shows the typical result obtained on two couples of films. After the thermal treatment in flowing Ar, any as-grown sample becomes superconducting and a convergence toward a specific crystalline cell is observed irrespective of the growth conditions. In fact, the superconducting films have the (004) peak in $2\theta_{SC} = 29.56°$ that corresponds to a value of the *c*-axis lattice parameter of 12.08 Å, in agreement with the literature [1, 9]. It is noteworthy that the as-grown films have completely different *c*-axis parameters: the *a*-type (004) peak is in a 2θ position higher than $2\theta_{SC}$, this denoting a shorter *c*-axis (i.e. 12.06Å), while the *b*-type (004) peak is in a 2θ position lower than $2\theta_{SC}$ corresponding to a longer *c*-axis (i.e. 12.13Å). Hence, the most oxygenated film behaves as typically shown in the literature, where the shortening of the *c*-axis parameter in the superconducting samples is reported [1]. The elongation of the same lattice parameter is the peculiarity of our *a*-type films: the inset of Fig. 1 shows an enlargement around the shift from blue to cyan curve in order to better appreciate the 2θ change position. The narrower full width at half maximum of the *a*-type superconducting peak respect to the *b*-type one is an indication of the best structural characteristic of the *a*-type samples.

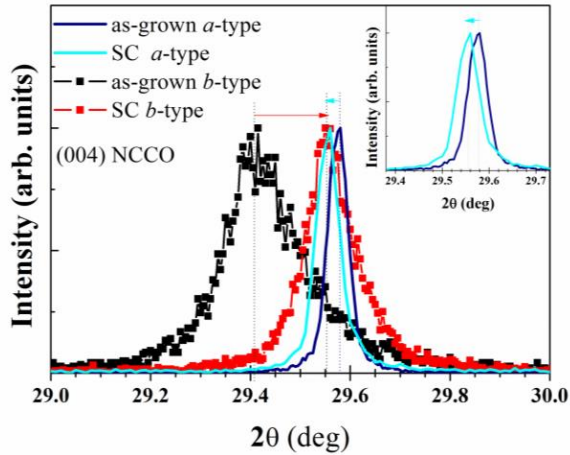

Fig. 1. θ-2θ measurements around the (004) reflection of the NCCO phase for a couple of *a*-type films and one of *b*-type samples. The blue and cyan lines refer to the *a*-type films reflections, while black and red squares and lines are used for the *b*-type films. The inset is an enlargement around the peak of the *a*-type samples, presented to better appreciate the 2θ change of position.

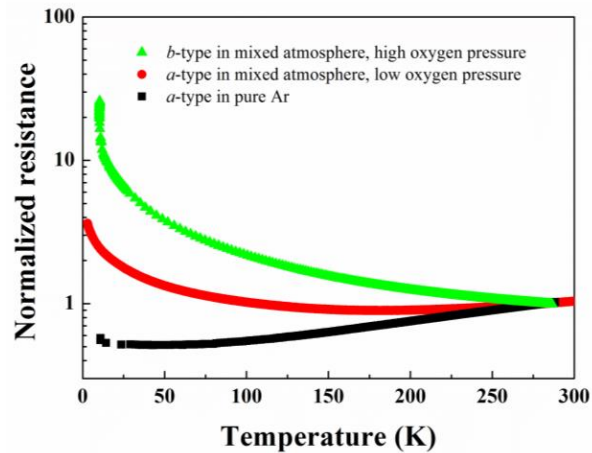

Fig. 2. R-T characteristics of as-grown film fabricated in different atmosphere conditions. The resistance is normalized to its value at T = 280 K.

The different oxygen content can be also linked with the resistance versus temperature (*R-T*) behavior. In particular, the *a*-type curve shows a minimum at a temperature $T_{min}$, that can change as a function of the growth condition [9], while the *b*-type one has a semiconducting-like behavior. In Fig. 2 typical *R-T* characteristics are reported. The minimum in the *R-T* curve is typically observed in the normal state of the under-doped cuprates but its interpretation is still under debate, particularly in the case of the electron-doped compounds [1]. The occurrence of a similar minimum in the *R-T* characteristics of the as-grown *a*-type film supports the idea that the oxygen role is crucial in the interpretation of the physics of the electron-doped cuprates, both in the normal and in the superconducting phase [14]. The minimum in the resistance curve is distinctive of a low Ce content, supposing an optimal content of oxygen. In our *a*-type film, by fixing the cerium doping x = 0.17, the change of the oxygen content arising from the growth condition of very low $O_2$ pressure reproduces the *R-T* curves obtained as a function of the Ce doping. Fig. 3 shows the resistance versus the magnetic field (*R-H*) curve obtained on an as-grown *a*-type film with $T_{min}$ = 50 K, at $T < T_{min}$: the negative MR (~ 10% at 1.6 K) is in agreement with the hypothesis of 2D weak localization induced by disorder as the origin of the insulating state at low temperature [17, 21].

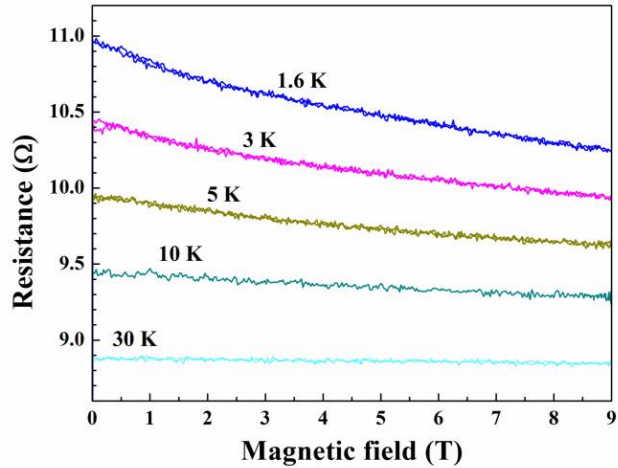

Fig.3. Negative magnetoresistance of an as-grown *a*-type film at $T < T_{min}$.

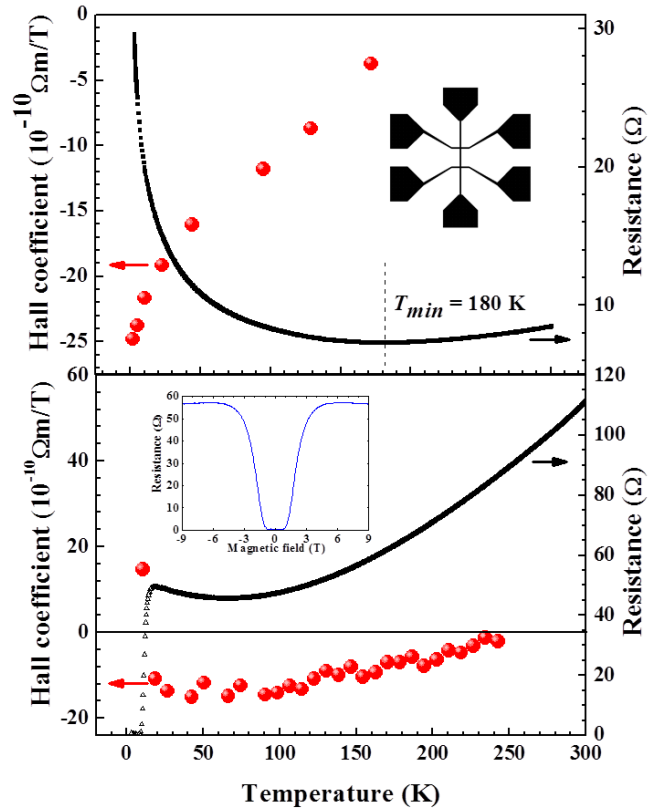

Fig.4. Temperature dependence of $R_H$ (red circles, left axis) and *R-T* (black squares, right axis), for an *a*-type as-grown (upper panel) and a superconducting sample (lower panel). The insets in the upper and in the lower panels show the Hall geometry adopted for our samples and the magnetoresistance of the superconducting film, respectively.

This interpretation is also supported by studies of the weak-localization effects based on electric noise spectroscopy. Indeed, it was demonstrated that the current dependence of the voltage noise shows an unusual linear behavior below $T_{min}$ as observed in copper thin films in the weak localized regime [22]. Finally, the temperature dependence of the Hall coefficient, $R_H$, was investigated for the *a*-type films, both in the normal and in the superconducting state (upper and lower panel of Fig. 4). The inset in the upper panel of Fig. 4 shows the Hall geometry adopted on our long strip samples having thickness, width and length equal to 100 nm, 100 μm and 1 mm, respectively. In the inset of the lower panel a negative MR at 1.6 K above the critical field of about 6 T is observed. Black squares are used for the *R-T* curve: the as-grown sample shows a $T_{min}$ of 180 K, the superconducting one has a critical temperature $T_c = 12$ K, slightly lower than the value reported in the literature [23], probably due to incomplete thermal treatments and/or to the possible degradation caused by the photolithography process. In Fig. 4 we also show the temperature dependence of the Hall coefficient, as calculated from the measurements of $R_{xy}$ versus $T$ at 9 T (red circles). The $R_H$ behavior that we observe is consistent with the data presented in the literature [14, 24, 25, 26]: $R_H$ for the as-grown sample is always negative, while the reduction process carries out the increase of $R_H$ and, consistently with the two-band model, the increase of the hole carriers mobility that induces the superconductivity [14]. The Ce doping only provides the necessary carriers for band filling, but this is not sufficient to get the superconductivity, as widely discussed in the literature [18, 27, 28]. In fact, the superconducting transition occurs only after thermal treatments that modify the O content in the samples. Hence, the change of the $R_H$ sign is strictly related to this O variation, irrespective of the cerium content. By fixing the Ce content in the superconducting state of NCCO samples, it is possible to measure $R_H > 0$ [25, 27] as well as $R_H < 0$ [24]. The comparison of these results with the data on $Pr_{1.83}Ce_{0.17}CuO_{4\pm\delta}$ [10, 14] and our measurements on $Nd_{1.83}Ce_{0.17}CuO_{4\pm\delta}$ thin films sheds light on the controversial results on the sign change of $R_H$. In fact, the study of our *a*-type and *b*-type films demonstrates that in both cases a reduction is required to induce the superconductivity, if the sample contains extra O, as in the case of the single crystals [25, 27], or the sample has O defects. Actually, the annealing procedure mainly allows the oxygen atoms mobility [6] and this in turn induces the sign change of $R_H$.

## IV. Conclusion

In conclusion, $Nd_{1.83}Ce_{0.17}CuO_{4\pm\delta}$ thin films have been grown in different $O_2$ pressure atmospheres. Two kinds of as-grown non-superconducting films were obtained, the *a*-type, deposited with low $O_2$ pressure, and the *b*-type, with higher $O_2$ pressure. It was pointed out that in the *a*-type film a $T_{min}$ is observed in the *R-T* curve, while the *b*-type exhibits a semiconducting-like behavior. In spite of the different *c*-axis lattice parameters and the different as-grown *R-T* behaviors, every sample becomes superconducting with the same *c*-axis parameter and $T_c$ ranging from 12 K and 24 K. The minimum in the *R-T* curve is distinctive of a low Ce content, under the hypothesis of an optimal content of O. In our *a*-type film, over-doped in cerium and under-doped in oxygen, the *R-T* curves appear similar to what reported in the literature for Ce content x < 0.15 [10]. The negative MR observed on our samples below $T_{min}$ is in agreement with the hypothesis of a low-temperature insulating behavior originated by 2D weak localization induced by disorder [17]. Our results on the evaluation of the Hall coefficient seem to indicate that the sign reversal of $R_H$ strictly depends on the oxygen atoms organization in the crystalline structure.

## Acknowledgment

Authors thank Dr. A. Vecchione for useful discussion on structural analysis.